\documentstyle[prl,aps]{revtex}

\begin{document}
\author{Jian-Qi Shen $^{1,}$$^{2}$ \footnote{E-mail address: jqshen@coer.zju.edu.cn}}
\address{$^{1}$  Centre for Optical
and Electromagnetic Research, State Key Laboratory of Modern
Optical Instrumentation, Zhejiang University,
Hangzhou Yuquan 310027, P. R. China\\
$^{2}$Zhejiang Institute of Modern Physics and Department of
Physics, Zhejiang University, Hangzhou 310027, P. R. China}
\date{\today }
\title{Tests of quantum-vacuum geometric phases via Casimir's effect}
\maketitle

\begin{abstract}
An experimentally feasible realization of testing quantum-vacuum
geometric phases of photons by using a gyrotropic-medium optical
fibre via Casimir's effect is proposed.

\\ \\
\end{abstract}
It was shown in quantum field theory that even in the vacuum, the
electromagnetic field does not really vanishes but rather
fluctuates. Vacuum thus has very large (even perhaps divergent)
zero-point energy density due to this quantum vacuum fluctuation,
which leads to Casimir's effect (where a weakly attractive force
exists between two parallel conducting metallic plates in a free
vacuum) and atomic spontaneous radiations (caused by the
interactions of excited atoms with vacuum zero-point
electromagnetic fields). In one of our published
papers\cite{1.,2.}, we suggested a new physically interesting
vacuum effect, {\it i.e.}, quantum-vacuum geometric phases of
photons propagating inside a noncoplanarly curved optical fibre.
Here we will propose an experimental realization of this geometric
phase at quantum-vacuum level by using gyrotropic-medium optical
fibres via Casimir's effect.

Berry first found that there exists a cyclic adiabatic geometric
phase (Berry's topological phase), in addition to the familiar
dynamical phase, of quantum mechanical wave function in a
time-dependent quantum adiabatic process\cite{3.,4.}. Geometric
phases is related close to the geometric nature of the pathway
along which quantum systems evolve and therefore possesses the
topological and global information on time evolution of quantum
systems, and hence attracts considerable attention of
investigators in various fields\cite{5.,6.,7.}, such as quantum
decoherence and topological quantum computation
recently\cite{8.,9.}. Chiao, Wu et al. proposed and treated in
detail a physical realization of Berry's cyclic adiabatic phase,
{\it i.e.}, the rotation of photon polarization planes of coiled
light in a helically curved optical fibre\cite{10.,11.,12.}. Based
on a second-quantized spin model\cite{12.,13.}, Gao and we have
previously studied the potential contribution of vacuum quantum
fluctuation to photon geometric phases in the noncoplanar fiber
and predicted the existence of quantum-vacuum geometric
phases\cite{1.,2.}.

For convenience, here we consider only the cyclic adiabatic
quantum-vacuum geometric phases of left- and right- handed
circularly polarized light in a helically coiled fibre, where the
direction of momentum (wave vector) of photons travelling along
the tangents of curved fibre is expressed by $(\sin
\theta\cos\varphi, \sin \theta\sin\varphi, \cos\theta)$ in the
spherical coordinate system. The geometric phases of left- and
right- handed circularly polarized photons are respectively
given\cite{1.}
\begin{equation}
 \phi_{L}=-\left(n_{L}+\frac{1}{2}\right)\cdot2\pi\left(1-\cos\theta\right),   \quad
 \phi_{R}=\left(n_{R}+\frac{1}{2}\right)\cdot2\pi\left(1-\cos\theta\right),              \label{1}
\end{equation}
where $n_{L}$ and $n_{R}$ denote the occupation numbers of left-
and right- handed (LRH) polarized photons, respectively. Note that
in a cyclic evolution process, $2\pi\left(1-\cos\theta\right)$ is
a solid angle subtended by a curve traced on a sphere, by the
direction of wave vector of light, at the centre, which manifests
the topological and global properties of time evolution of photon
wave function in the fibre. It follows from expressions (\ref{1})
that the quantum-vacuum geometric phases resulting from quantum
vacuum fluctuation are $\phi^{\rm
vac}_{\pm}=\pm\left(\frac{1}{2}\right)\cdot2\pi\left(1-\cos\theta\right)$
with the sign $\pm$ corresponding to the right- and left- handed
polarized light, respectively. Note that here the quantum-vacuum
phase of left-handed polarized light is different from that of
right-handed polarized light only by a minus sign. This,
therefore, means that the LRH quantum-vacuum geometric phases may
be cancelled by each other and in consequence, unfortunately,
their presence cannot be easily detected experimentally, since the
vacuum zero-point LRH polarized electromagnetic fields are present
often accompanied by one another in conventional optical fibres.

So, in the previous fibre experiments\cite{11.,12.}, the observed
total geometric phases are
$\phi_{L}+\phi_{R}=\left(\phi_{R}-\phi_{L}\right)\cdot2\pi\left(1-\cos\theta\right)$,
which is not associated with vacuum fluctuation. Here we suggest
strongly that the quantum-vacuum geometric phases also deserved
experimental investigation, since it has an important connection
with the topological properties of time evolution of quantum
vacuum fluctuation. However, the problem we encounter now is: can
we truly extract the non-vanishing vacuum geometric phase of left-
or right- handed polarized light from the cancelled (and hence
vanishing) total vacuum geometric phases? More recently,  we
propose an idea on experimental realization of this quantum-vacuum
geometric phases in the fibre. We think the peculiar wave
propagation properties in gyrotropic media may be applicable to
the problem mentioned above.

Gyrotropic media is such electromagnetic materials where both the
electric permittivity and the magnetic permeability are tensors,
which can be respectively written as
\begin{equation}
(\epsilon)_{ik}=\left(\begin{array}{cccc}
\epsilon_{1}  & i\epsilon_{2} & 0 \\
-i\epsilon_{2} &   \epsilon_{1} & 0  \\
 0 &  0 &  \epsilon_{3}
 \end{array}
 \right),                 \qquad          (\mu)_{ik}=\left(\begin{array}{cccc}
\mu_{1}  & i\mu_{2} & 0 \\
-i\mu_{2} &   \mu_{1} & 0  \\
 0 &  0 &  \mu_{3}
 \end{array}
 \right) .              \label{2}
\end{equation}
Assuming that the direction of wave vector of electromagnetic wave
is parallel to the third component of the Cartesian coordinate
system, the optical refractive indices squared of gyrotropic media
corresponding to the two directions of polarization vectors are of
the form
$n^{2}_{\pm}=(\epsilon_{1}\pm\epsilon_{2})(\mu_{1}\pm\mu_{2})$. It
is well known that in a finitely large space ({\it e.g.}, the
space between two parallel metallic plates, which is the main
equipment of Casimir's effect experiment), the vacuum-fluctuation
electromagnetic field alters its mode structures, namely, the
zero-point field with wave vector $k$ less than
$\sim\left(\frac{\pi}{a}\right)$ does not exist in the space with
a finite scale length $a$. In some certain gyrotropic media, if
the electromagnetic parameters $\epsilon_{1}$, $\epsilon_{2}$ and
$\mu_{1}$, $\mu_{2}$ in expressions (\ref{2}) for electric
permittivity and magnetic permeability tensors can be chosen to be
$\epsilon_{1}\simeq\epsilon_{2}$ and $\mu_{1}\simeq\mu_{2}$, then
the optical refractive index $n_{-}$ of left-handed polarized
light is very small (or tending to zero) and hence the wave vector
$k_{-}\simeq 0$ , since in media $k_{-}$ is proportional to
$n_{-}$. Thus the left-handed polarized zero-point field inside
these gyrotropic media is absent if, for example, the media are in
a finitely large space, and consequently the only retained
quantum-vacuum geometric phase is that of right-handed circularly
polarized light. In order to perform the detection of this
geometric phase, we should make use of the optical fibre that is
fabricated from the above gyrotropic media, and the devices used
in Tomita-Chiao fibre experiments should be placed in a sealed
metallic chamber or cell, where the zero-point circularly
polarized field with lower wave vector does not exist. Thus the
measurement of quantum-vacuum geometric phases of photons may be
achievable by means of this somewhat ingenious scheme.

Although the infinite vacuum energy in conventional {\it
time-independent} quantum field theories is harmless and easily
removed theoretically by normal-order procedure, here for a {\it
time-dependent} quantized-field system, we think the existence of
quantum-vacuum geometric phases indicates that zero-point fields
of vacuum will also participate in the time evolution process and
perhaps can be no longer regarded merely as an inactive onlooker
in {\it time-dependent} quantum field theories such as field
theory in curved space-time, {\it e.g.}, in time-dependent
gravitational backgrounds and expanding universe. In order to
investigate this fundamental problem of quantum field theory, we
hope the vacuum effect presented here would be tested
experimentally in the near future. We also hope to see anyone else
putting forward some new more clever suggestions of detecting the
interesting quantum-vacuum geometric phases.

\textbf{Acknowledgements}  This project is supported partially by
the National Natural Science Foundation of China under the project
No. $90101024$.


\begin{references}
\bibitem{1.}  Shen, J. Q. and Ma, L. H. Phys. Lett. A 308, 355-363
(2003). A supplement to this paper is: Shen, J. Q.
quant-ph/0304172 at $<$xxx.lanl.gov$>$ (2003).

\bibitem{2.}  Gao, X. C. Chin. Phys. Lett. 19, 613-616 (2002).

\bibitem{3.}  Berry, M. V. Proc. R. Soc. London, Ser. A 392, 45-57 (1984).

\bibitem{4.}  Berry, M. V. Nature 326, 277-278 (1987).

\bibitem{5.} Falci, G. et al. Nature 407, 355-358 (2000).

\bibitem{6.}  Taguchi, Y. et al. Science 291, 2573-2576 (2001).

\bibitem{7.}  Jones, J. A. Nature 403, 869-871 (2000).

\bibitem{8.} Zhu, S. L. and Wang, Z. D. Phys. Rev. Lett. 89, 097902 (2002).

\bibitem{9.} Wang, X. B. and Keiji, M. Phys. Rev. Lett. 87, 097901 (2001).

\bibitem{10.} Chiao, R.Y. and Wu, Y. S. Phys. Rev. Lett. 57, 933-936
(1986).

\bibitem{11.} Tomita, A. and Chiao, R. Y. Phys. Rev. Lett. 57, 937-940
(1986).

\bibitem{12.} Robinson, L. Science 234, 424-426 (1986).

\bibitem{13.} Shen, J. Q. and Zhu, H. Y. Ann. Phys. (Leipzig) 12, 131-145
(2003).


\end{references}
\end{document}